\documentstyle[12pt]{article}

\catcode`\@=11
\long\def\@makefntext#1{
\protect\noindent \hbox to 3.2pt {\hskip-.9pt
$^{{\ninerm\@thefnmark}}$\hfil}#1\hfill}		

\def\@makefnmark{\hbox to 0pt{$^{\@thefnmark}$\hss}}  

\def\ps@myheadings{\let\@mkboth\@gobbletwo
\def\@oddhead{\hbox{}
\rightmark\hfil\ninerm\thepage}
\def\@oddfoot{}\def\@evenhead{\ninerm\thepage\hfil
\leftmark\hbox{}}\def\@evenfoot{}
\def\sectionmark##1{}\def\subsectionmark##1{}}

\renewcommand{\thefootnote}{\fnsymbol{footnote}}

\newcounter{sectionc}\newcounter{subsectionc}\newcounter{subsubsectionc}
\renewcommand{\section}[1] {\vspace*{0.6cm}\addtocounter{sectionc}{1}
\setcounter{subsectionc}{0}\setcounter{subsubsectionc}{0}\noindent
	{\normalsize\bf\thesectionc. #1}\par\vspace*{0.4cm}}
\renewcommand{\subsection}[1] {\vspace*{0.6cm}\addtocounter{subsectionc}{1}
	\setcounter{subsubsectionc}{0}\noindent
	{\normalsize\it\thesectionc.\thesubsectionc. #1}\par\vspace*{0.4cm}}
\renewcommand{\subsubsection}[1]
{\vspace*{0.6cm}\addtocounter{subsubsectionc}{1}
	\noindent {\normalsize\rm\thesectionc.\thesubsectionc.\thesubsubsectionc.
	#1}\par\vspace*{0.4cm}}

\newcounter{appendixc}
\newcounter{subappendixc}[appendixc]
\newcounter{subsubappendixc}[subappendixc]

\renewcommand{\appendix}[1] {\vspace*{0.6cm}
        \refstepcounter{appendixc}
        \setcounter{figure}{0}
        \setcounter{table}{0}
        \setcounter{equation}{0}
        \renewcommand{\thefigure}{\Alph{appendixc}.\arabic{figure}}
        \renewcommand{\thetable}{\Alph{appendixc}.\arabic{table}}
        \renewcommand{\theappendixc}{\Alph{appendixc}}
        \renewcommand{\theequation}{\Alph{appendixc}.\arabic{equation}}
        \noindent{\bf Appendix \theappendixc #1}\par\vspace*{0.4cm}}

\def\abstracts#1{{
	\centering{\begin{minipage}{12.2truecm}\footnotesize\baselineskip=12pt\noindent
	\centerline{\footnotesize ABSTRACT}\vspace*{0.3cm}
	\parindent=0pt #1
	\end{minipage}}\par}}

\renewenvironment{thebibliography}[1]
	{\begin{list}{\arabic{enumi}.}
	{\usecounter{enumi}\setlength{\parsep}{0pt}
\setlength{\leftmargin 1.25cm}{\rightmargin 0pt}
	 \setlength{\itemsep}{0pt} \settowidth
	{\labelwidth}{#1.}\sloppy}}{\end{list}}

\newcounter{itemlistc}
\newcounter{romanlistc}
\newcounter{alphlistc}
\newcounter{arabiclistc}

\newcommand{\fcaption}[1]{
        \refstepcounter{figure}
        \setbox\@tempboxa = \hbox{\footnotesize Fig.~\thefigure. #1}
        \ifdim \wd\@tempboxa > 6in
           {\begin{center}
        \parbox{6in}{\footnotesize\baselineskip=12pt Fig.~\thefigure. #1}
            \end{center}}
        \else
             {\begin{center}
             {\footnotesize Fig.~\thefigure. #1}
              \end{center}}
        \fi}

\newcommand{\tcaption}[1]{
        \refstepcounter{table}
        \setbox\@tempboxa = \hbox{\footnotesize Table~\thetable. #1}
        \ifdim \wd\@tempboxa > 6in
           {\begin{center}
        \parbox{6in}{\footnotesize\baselineskip=12pt Table~\thetable. #1}
            \end{center}}
        \else
             {\begin{center}
             {\footnotesize Table~\thetable. #1}
              \end{center}}
        \fi}

\def\@citex[#1]#2{\if@filesw\immediate\write\@auxout
	{\string\citation{#2}}\fi
\def\@citea{}\@cite{\@for\@citeb:=#2\do
	{\@citea\def\@citea{,}\@ifundefined
	{b@\@citeb}{{\bf ?}\@warning
	{Citation `\@citeb' on page \thepage \space undefined}}
	{\csname b@\@citeb\endcsname}}}{#1}}

\newif\if@cghi
\def\cite{\@cghitrue\@ifnextchar [{\@tempswatrue
	\@citex}{\@tempswafalse\@citex[]}}
\def\citelow{\@cghifalse\@ifnextchar [{\@tempswatrue
	\@citex}{\@tempswafalse\@citex[]}}
\def\@cite#1#2{{$\null^{#1}$\if@tempswa\typeout
	{IJCGA warning: optional citation argument
	ignored: `#2'} \fi}}

 1
 1
 1

\font\ninerm=cmr9

\begin{document}

\centerline{\normalsize\bf Roper excitation in $\vec{p}+\alpha \rightarrow 
\vec{p}+\alpha+X$ reactions }
\baselineskip=16pt

\vspace*{0.6cm}
\centerline{\footnotesize S. Hirenzaki }
\baselineskip=16pt
\centerline{\footnotesize\it  Department of Physics, 
Nara Women's University, Nara 630-8506, Japan}
\vspace*{0.6cm}
\baselineskip=12pt
\centerline{\footnotesize A. D. Bacher and S. E. Vigdor}
\baselineskip=12pt
\centerline{\footnotesize\it  Dept. of Physics, Indiana University, 
Bloomington, Indiana 47405, U.S.A.}
\baselineskip=12pt

\vspace*{0.9cm}
\abstracts{We calculate differential cross sections and the spin transfer 
coefficient $D_{nn}$ 
in the  $\vec{p}+\alpha \rightarrow \vec{p}+\alpha+\pi^0$ reaction for 
proton bombarding energies from 1 to 10 GeV and $\pi^0 - p$ invariant masses
spanning the region of the N$^*$(1440) Roper resonance.  Two processes --
$\Delta$ excitation in the $\alpha$-particle and Roper excitation in the
proton -- are included in an effective reaction model 
which was shown previously to reproduce existing inclusive spectra.
The present calculations demonstrate that these two contributions can
be clearly distinguished via $D_{nn}$, even under kinematic conditions
where cross sections alone exhibit no clear peak structure due to the
excitation of the Roper. }

\vspace*{0.8cm}

\noindent
[PACS: 14.20.Gk, 25.40.-h, 13.75.-n]

\vspace*{0.8cm}
\normalsize\baselineskip=15pt
\setcounter{footnote}{0}
\renewcommand{\thefootnote}{\alph{footnote}}
\section{Introduction}

An important goal of theoretical approaches to non-perturbative QCD is
to reproduce the spectrum and properties of nucleon resonances in terms of 
quark and gluon constituents.  The excited baryons with the same quantum 
numbers as the nucleon -- e.g.,
the N$^*$(1440) Roper resonance and the N$^*$(1710) -- are particularly
poorly understood at present.  It has been difficult to understand in
models why an excited configuration of three constituent quarks with the
same quantum numbers as the nucleon would lie as low in mass as 1440 MeV
\cite{Stoler93}.  This problem has opened the door to speculative alternative
interpretations of the structure of the Roper resonance, e.g., involving
collective excitations of the nucleon \cite{Bijker96} or hybrid states with 
more valence constituents than three quarks \cite{Li92}.  Tests of such 
structure models have been impeded by experimental difficulties in exciting 
the Roper selectively.

Recent experiments at the Laboratoire National Saturne \cite{Morsch92}
have provided encouraging signs that the $(\alpha,\alpha')$ reaction on 
the proton may provide a method for such selective excitation.  Two
distinct peaks observed in small-angle inclusive $\alpha$-particle inelastic
scattering spectra at $T_\alpha = 4.2 ~GeV$ were interpreted as arising,
respectively, from $\Delta$ excitation in the $\alpha$ projectile (DEP) 
and Roper excitation in the proton target \cite{Morsch92}. A subsequent 
theoretical analysis \cite{Hirenzaki96} demonstrated that this picture 
is indeed qualitatively consistent with the measured inclusive spectra.  
The above two mechanisms, illustrated in Fig. 1, were shown \cite{Hirenzaki96} 
to dominate over other possible mechanisms, such as Roper excitation in 
the projectile or excitation of two $\Delta$-particles.  However, it was 
also found that the interference between the two mechanisms in Fig. 1 is 
appreciable, and necessary to consider for a quantitative account 
of the data.  In other work \cite{Napolitano97}, the identification of
the second observed peak in $\alpha -$p inelastic scattering as arising 
entirely from the Roper resonance has been called into question, on the 
basis of multipole decompositions of a high statistics sample of events 
from the $K^- p \rightarrow K^- p \pi^+ \pi^-$ reaction.

It is thus interesting to consider, within the framework of the same 
reaction model \cite{Hirenzaki96}, whether other experiments in 
the p--$\alpha$ system may exhibit enhanced sensitivity to the Roper
excitation amplitudes.  For example, it was subsequently predicted 
\cite{Hirenzaki96-2} that the signal for Roper excitation should be 
strongly enhanced
with respect to the DEP background in p($\alpha,\alpha^\prime$) reactions
by raising the $\alpha$-particle bombarding energy to 10-15 GeV.  In the 
present paper,
we demonstrate the value of {\em polarization transfer} measurements
in exclusive $\vec{p}+\alpha \rightarrow \vec{p}+\alpha+X$ reactions
for distinguishing the Roper (isoscalar, non-spin-flip) excitation 
from $\Delta$ (isovector, spin-flip) excitation.  The utility of 
polarization transfer measurements for distinguishing analogous 
{\em nuclear} 
transitions has been clearly demonstrated in medium-energy proton-nucleus 
reaction studies \cite{Moss82}.  

In the present case, if the reaction proceeds 
through an intermediate $\Delta$, we expect a negative value 
$D_{nn} < 0$ for the transfer of normal 
polarization from the incident proton to the final-state proton 
\cite{Indiana}, in analogy with the results for Gamow-Teller transitions 
in nuclei with A(${\vec p},{\vec n}$) reactions at moderate momentum 
transfer \cite{Taddeucci84}.  
In contrast, the simple spin structure for the direct 
excitation of the Roper by an $\alpha$-particle --
$0^+ + \frac{1}{2}^+ \rightarrow 0^+ + \frac{1}{2}^+$ --
requires $D_{nn} = 1$ by parity 
conservation \cite{Csonka66,Csonka66-2,Weiden70}.  
Furthermore, for the Roper decay mode N$^* \rightarrow$ N + $\pi$,
the polarization of the Roper is completely 
transferred to its daughter proton when the proton is emitted along 
the Roper polarization axis in the Roper rest frame.  Thus, for a 
restricted region of phase space in a coincidence measurement
$\vec{p}+\alpha \rightarrow \vec{p}+\alpha+X$, one can expect to 
distinguish the Roper contribution from the $\Delta$ contribution 
by observing $D_{nn}$, even if one does not see
a clear peak in cross section spectra.  These ideas have been 
described previously\cite{Indiana}, but only in a qualitative manner.  

In the present work, we carry out quantitative calculations for
differential cross sections and $D_{nn}$ in the exclusive 
$\vec{p}+\alpha \rightarrow \vec{p}+\alpha+\pi^0$ reaction at several
bombarding energies, including both mechanisms in Fig.~1.  In our
model, we include proton-$\alpha$ distortions using a spin-independent
eikonal approximation.  We expect this model to be reasonably good for
predicting cross sections and $D_{nn}$, since the $D_{nn}$-value for
Roper excitation is fixed by parity conservation, independent of distortions
and other details of the production mechanism.  On the other hand, this
simple treatment of distortions may be inadequate for 
other, less robust spin observables, such as the analyzing power $A_y$.  

The paper is organized as follows.  Section 2 describes the theoretical
model for the $\vec{p}+\alpha \rightarrow \vec{p}+\alpha+X$ reactions.  
Section 3 presents the numerical results of the reaction model.  Section 4
summarizes the results and indicates possible applications of this technique
to other nucleon excitations.   

\section{Model for the $\vec{p}+\alpha \rightarrow 
\vec{p}+\alpha+X$ reactions}

We use the same model developed in 
Refs. 5, 7, 14 and refer the reader to these references for details.
We include the two dominant processes shown in Fig.~1 -- $\Delta$ 
excitation in the $\alpha$-particle and Roper excitation in the proton --
which are necessary to reproduce the inclusive cross section spectra from
Ref.~4.  We can write the amplitudes as: 

\begin{equation}
 - i T^{\Delta}_{m'm} = - \frac{16}{9} F_{\alpha} \left( \frac{f^*}{\mu} \right)^2 
  \left( \frac{f}{\mu} \right) G_{\Delta} 
  \sqrt{ \frac{-q^2}{\vec{q_{\Delta}}^2} } 
  [ (V_{l'} - V_{t'} ) (\vec{p}_{\Delta} \cdot 
 \hat{q}_{\Delta}) \hat{q}_{N} + V_{t'} \vec{p}_{\Delta} ] \cdot 
 <m' | \vec{\sigma} | m >  ,
\end{equation}
 
\noindent
and 

\begin{equation}
 - i T^{*}_{m'm} = - 4 F_{\alpha}  \left( \frac{f'}{\mu} \right) G_* g_{\sigma 
 N N^*} D_{\sigma} F_{\sigma}^2 g_{\sigma NN} \vec{p_*} \cdot <m' | \vec{\sigma} | m > .
\end{equation}

\noindent 
where $G_{\Delta}$ and $G_*$ are the propagators of the $\Delta$ and Roper 
resonances, $D_{\sigma}$ is the propagator of the $\sigma$ meson, 
$F_{\alpha}$ is the $^4$He nuclear form factor, $\mu$ is the pion mass, 
and $F_{\sigma}$ is the 
$\sigma NN$ vertex form factor.
$V_{l'}$ and $V_{t'}$ stand for the longitudinal and transverse parts 
of the $NN \rightarrow N\Delta$ 
effective interaction which includes $\pi$, 
$\rho$, and $g'$ contributions.  
The $f$'s and $g$'s in Eqs. (1) and (2) are coupling constants.  
In particular, $f'$ is determined to 
reproduce the decay width of the $N^*(1440) \rightarrow \pi N$ channel.  
All details, including parameter values, are given in Refs. 5, 14.
In Eqs. (1) and (2), the subscripts on momenta, $\Delta$, $N$, and $*$, 
indicate the coordinate system where the momenta are to be evaluated:  
the $\Delta$ rest frame, the initial proton rest frame, and the Roper 
rest frame, respectively.  The magnetic quantum numbers $m$ and $m'$ for
initial and final protons refer to a spin quantization axis perpendicular
to the reaction plane formed by the beam and outgoing proton or N$^*$
directions.  

In the amplitudes we include only $p+\alpha+\pi^0$ as the final 
state.  In $p-\alpha$ coincidence experiments, the missing mass of the 
$\pi^0$ can be reconstructed to eliminate contributions from $2\pi$ 
decay channels of the Roper resonance.  However, their neglect in the 
calculations reported here for inclusive spectra is expected to yield
an underestimate of the cross section for the Roper process in the higher 
excitation energy region of the inclusive spectra.  The $2\pi$ decay 
channels mainly 
contribute to the inclusive cross section at higher excitation energy 
because of the larger available phase space.  They will make the Roper 
contribution to inclusive spectra broader than shown here, especially 
at the higher incident energies. \cite{Hirenzaki96-2}
 
The nuclear form factor $F_\alpha$ contained in Eqs. (1) and (2) is 
defined as

\begin{displaymath}
  F_{\alpha}(\vec{k})  = \frac{1}{4} \int d^3 r \rho_{\alpha} (\vec{r})
               exp \left[ - \frac{1}{2} \int _{-\infty} ^{\infty} \sigma_{NN}
               \rho_{\alpha} (\vec{b}, z') dz' \right]
               e^{i \vec{k} \cdot \vec{r} }
\end{displaymath}

\begin{equation}
               \times exp \left[ - \frac{i}{2} \int _{0} ^{\infty}
               \frac{1}{p_{\pi}} \Pi (p_{\pi}, \rho_{\alpha} (\vec{r'} ) ) d \ell
               \right]  ,
\end{equation}

\noindent
where
\begin{displaymath}
   \vec{r'} = \vec{r} + \frac{\vec{p_{\pi}}}{\vert \vec{p_{\pi}} \vert} \ell ,
\end{displaymath}

\begin{equation}
  \vec{k} = \vec{p_{\alpha}} - \vec{p_{\alpha'}}  ,
\end{equation}

\noindent
and $\vec{b}$ is the impact parameter.  
We write $F_\alpha (\vec{k}) $ normalized to unity at $\vec{k}=0$ and in 
the absence of distortion, as is usually done.  The momenta $ 
\vec{p_{\alpha}}, \vec{p_{\alpha'}}, \vec{p_{\pi}}$ appearing in Eqs. 
(3) and (4) are evaluated in the frame where the initial $\alpha$-particle 
is at rest.  In Eq. (3), $\rho_{\alpha} (\vec{r} )$ is a harmonic  
oscillator density distribution of $^4$He, $\sigma_{NN}$ is the 
nucleon-nucleon total cross section and $\Pi (p_{\pi}, \rho ) / 2 
\omega_\pi$ is the pion nuclear optical potential with the relativistic 
pion energy $\omega_\pi$.  In this definition of the 
$F_{\alpha}(\vec{k})$, we apply the eikonal approximation, which is known
to be a good approximation at intermediate energies, to evaluate
distortion effects.  In addition, we neglect nonlocality due to
meson exchange, and also the propagation of $\Delta$ and N$^*$, because
of their large widths and prompt decay.

The observed inclusive cross sections led the authors of Ref.~4 
to interpret the Roper resonance as the E0 monopole 
excitation of the nucleon.  
However, in our theoretical model, the monotonic decrease of the observed 
angular distribution\cite{Morsch92} is mostly a consequence of the $^4$He 
form factor and not an intrinsic property of the Roper resonance.  Our 
calculated results reproduce the trend of all of the experimental results 
quite well \cite{Hirenzaki96} 
without treating the Roper as the monopole excitation of the nucleon.  
We think that the limited information in the data obtained so far does 
not allow one to extract such precise information on the structure of the Roper.  

Using the amplitudes shown in Eqs. (1) and (2), the coincidence
cross section can be written as 

\begin{equation}
 \frac{d \sigma}{dE_{\alpha'} d \Omega_{\alpha'} d \Omega_{p'} }
 = \frac{p_{\alpha'}}{ (2 \pi)^5} \frac{M^2_{\alpha} M^2}{\lambda^{1/2} 
 (s,M^2,M^2_{\alpha})} \frac{p'^2}{p' \omega_{\pi} + E' (p' - p_{\pi N} 
 cos \theta _2 )} 
 \bar{\sum_{m}} \sum_{m'} | T^{\Delta}_{m'm} + T^{*}_{m'm} | ^2 .
\end{equation}

\noindent
where $M$ is the nucleon mass, $M_{\alpha}$ is the mass of the $^4$He, and 
$\lambda ( \cdot \cdot \cdot )$ is the Kallen function defined as; 

\begin{displaymath}
\lambda (a,b,c) = a^2 + b^2+ c^2 -2ab - 2bc - 2ca.
\end{displaymath}

\noindent
All kinematical variables are evaluated in the laboratory frame 
and defined in Fig. 2.   

The normal spin transfer coefficient $D_{nn}$ is 
defined as; 

\begin{equation}
 D_{nn} = \frac{(d\sigma_{uu}+d\sigma_{dd})-(d\sigma_{ud}+d\sigma_{du})}
 {(d\sigma_{uu}+d\sigma_{dd})+(d\sigma_{ud}+d\sigma_{du})}
\end{equation}

\noindent
where the indices, $u$ and $d$, indicate the up and down spin state of the 
proton in the initial and final states.  Here, the cross sections 
$d\sigma_{m'm}$ are defined 
by Eq. (5) without taking the spin sum and average.  

\section{Numerical Results}

We first calculate cross sections for the inclusive reaction $p + \alpha 
\rightarrow \alpha + X$, which is the same inclusive reaction considered in 
Ref.~5, except for altered kinematics. In the present 
case, the proton is the projectile and the recoiling $\alpha$-particle is 
observed in the final state.  We use 
the same $T$ matrix defined in Section 2 and the same phase 
factors as in Ref. 5.  Since we may also have the $ n + 
\alpha + \pi^+$ final state in the inclusive reaction, we 
have multiplied by an additional isospin factor of 3 the cross sections 
which are obtained using the $T$ matrix from Section 2.    We calculate 
the inclusive cross section $d\sigma / dE_{\alpha'} d\Omega _{\alpha'}$ as 
a function of $T_{\alpha'}$ at different $\alpha '$ angles $\theta_1$ 
(see Fig. 2. for the definition of $\theta_1$).  The calculated results 
are shown in Figs. 3 and 4 for incident proton energies of $1 ~GeV$ and 
$10 ~GeV$.  We also show the contributions to the inclusive cross section 
from the Roper excitation process alone.  

In Fig. 3, we show the calculated results for $T_p=1 ~GeV$, which 
corresponds to $T_{\alpha}=4 ~GeV$ in the inverse kinematics of the 
Saturne experiment \cite{Morsch92}.  It is interesting to compare our 
present results with the measurements from Saturne.  The shape of the 
energy spectrum at $\theta_{1} = 20^\circ$ 
in Fig. 3(b) strongly resembles that observed in the inverse kinematics 
\cite{Hirenzaki96}.   We find, however, that the angular dependence 
of the cross section is much milder in the present case than for the case 
of an $\alpha$-particle projectile: in going from $\theta_1 = 0^\circ$ to
$60^\circ$, the cross section decreases by only about a factor of three.  
This mild angular dependence is due to the behavior 
of the $\alpha$-$\alpha'$ transition form factor $F_{\alpha}$ in Eqs. (1) 
and (2).  We evaluate the form factor using the momentum transfer for 
the $\alpha$-particle in the initial $\alpha$ rest frame.  In the present 
kinematics, the momentum transfer does not depend on the angle $\theta_1$,
but only on the energy $E_{\alpha'}$.   
Thus, $F_{\alpha}$, which caused the steep angular dependence observed in the 
case of inverse kinematics, does not produce an angular dependence of the energy 
spectrum calculated here.  The observed dependence of the spectra in Figs. 3 
and 4 on $\theta_{1}$ arises instead from kinematic effects described below.  

In Fig. 3 we can see the cross sections from the Roper process alone at 
different values of $\theta_1$ for $T_p=1 ~GeV$.  As $\theta_1$ increases, 
the Roper peak 
moves to larger $T_{\alpha'}$ and becomes weaker and broader.  
This behavior reflects changes in the invariant mass of the 
final $\pi N$ system.  Since the invariant mass changes more slowly as a 
function of $T_{\alpha'}$ for larger angles, the peak position moves to 
larger $T_{\alpha'}$ and the peak is broader 
when we plot cross sections as a function of 
$T_{\alpha'}$.  For larger $T_{\alpha'}$, the transition form 
factor $F_{\alpha}$ makes the Roper peak weaker.  Furthermore, at $\theta_1 = 
40^\circ$ and $60^\circ$ with $T_p = 1 ~GeV$, the invariant mass cannot 
reach 1440 MeV, so that the Roper contributions are much smaller than 
at more forward $\alpha$-particle angles.

The contribution from the $\Delta$ process has a different angular 
dependence, as can be seen in Fig. 3, since the invariant mass of the 
$\Delta$ system is determined in a different way (see Ref. 
14).  Nonetheless, the $\Delta$ peak also moves to larger $T_{\alpha'}$ for 
larger $\theta_1$, and decreases in strength as a result of $F_{\alpha}$.  

For higher $T_p$ (see Fig.~4), the Roper contribution is larger than 
the $\Delta$ contribution, since the Roper peak moves to smaller 
$T_{\alpha'}$, where $F_{\alpha}$ is larger.  This is also the case in 
inverse kinematics, as reported in Ref.~7.  At the same 
time, the Roper peak is sharper because the invariant 
mass changes more rapidly as a function of $T_{\alpha'}$.  In the 
present case, however, the Roper and $\Delta$ peaks strongly overlap for 
higher $T_p$, and cannot be distinguished in inclusive spectra alone. 

The angular dependence of the cross section for both the Roper and the 
$\Delta$ excitation processes in Fig.~4 is much flatter than at lower
$T_p$, because of the $p_{\alpha'}$ included in the phase space factor 
of the cross section.  The increase of $p_{\alpha'}$ at larger $\theta_1$ 
overcomes the effect of  
$F_{\alpha}$ in this narrow energy range close to $p_{\alpha'}=0$, 
making the cross section larger.  

The inclusive spectra shown in Figs.~3 and 4 indicate that the 
$\alpha$-particle recoil energy is quite small in the laboratory frame, 
and that good energy resolution is needed 
to select the portion dominated by Roper excitation.  This fact favors 
the use of a thin, windowless gaseous $^4$He target in an experiment.  
The use of a storage 
ring and internal target environment, as proposed in Ref.~9, 
seems to be most suitable to obtain sufficient luminosity.

Before presenting the numerical results for {\em exclusive} reactions, 
we need to 
clarify the kinematic configurations in which we calculate the exclusive 
cross sections.  As described in section 1, we are interested in the 
restricted phase space in the final state where the spin transfer 
coefficient $D_{nn}$ of the Roper process is equal to one.  
In the present reaction, the energy and momentum of the Roper are
determined uniquely for each 
final $\vec{p}_{\alpha'} $.  Furthermore, the normal polarization of
the proton beam is transferred completely to the produced Roper.
When the Roper decays into the $\pi 
+ p$ system, we can determine the desired momenta and energies of the 
$\pi$ and $p$ uniquely by imposing the additional condition that the 
proton be emitted along the polarization axis of the Roper within the 
Roper rest frame.  This condition guarantees full transfer of the Roper 
polarization to its daughter proton.  The final proton energy and 
emission angles in the laboratory frame (see Fig.~2) are then obtained
by a Lorentz transformation from the Roper rest frame to the 
laboratory frame.  In this restricted kinematic configuration, we always 
get $D_{nn} = 1$ for the Roper contribution.  As an example, Fig. 5 shows, 
for the case of $T_p=2 ~GeV$, the final proton emission angles 
($\theta_2,~\theta_3$) and kinetic energies as a function of $T_{\alpha'}$ 
for several values of $\theta_1$.  All of our 
results for the exclusive reaction are obtained in this kinematic 
condition. Thus, 
the final proton energy and angles in the laboratory frame vary with 
those of the final $\alpha$, so as to satisfy the conditions described 
above.  Note, however, from Fig.~5 that the final proton remains less
than 20$^\circ$ out of plane ($\theta_2 < 20^\circ$) over the entire
range of interest, so that its polarization is always predominantly
transverse to its motion in the laboratory frame.  Furthermore, the decay 
proton energies in the lab frame are in a range near that where 
high figure-of-merit proton polarimeters have already been 
developed at LAMPF \cite{LAMPF}.

Experiments will, of course, average over finite angular and energy 
acceptances for the decay proton.  Thus, we have also considered
final protons emitted at non-zero angles from the polarization axis 
in the Roper rest frame.  We find, for example, that in the $T_p=2 ~GeV$, 
$\theta_1=20^\circ$ case, one maintains $D_{nn} > 0.95 $ for the 
Roper process at its peak if decay protons are detected over a $\pm 0.5^\circ$
angular and $\pm 25$ MeV energy range in the laboratory, centered around
the optimum values.

In Figs. 6 and 7 we show the calculated exclusive reaction cross 
sections and the spin transfer 
coefficient [defined in Eqs. (5) and (6)], for $T_p = 1 ~GeV$.  
Since the phase space factor of Eq. (5) 
diverges at the threshold for one pion 
production, the total cross sections are larger at smaller $T_{\alpha'}$.  At 
the threshold, $D_{nn}$ for the $\Delta$ process is $-1$ since both the  
proton and the pion in the final state are in the scattering plane, so that
the momentum transfer to the nucleon is perpendicular to the spin 
polarization.  $D_{nn}$ for the Roper process is always 1 in the 
kinematic configuration described above.  
The $D_{nn}$ associated with the interference between the two contributions
is also always 1, because the interference makes a finite 
contribution only when the amplitude for the Roper process is non-zero.
In Fig. 6, where $\theta_1 = 20^\circ$, we see that the calculated 
cross section does not exhibit a clear peak due to the Roper contribution,
but rather only a shoulder.  Nonetheless, in the spin transfer coefficient 
one sees a clear 
indication of the Roper excitation process:   $D_{nn}$ clearly changes from 
negative to positive $ ( \sim 1) $ in the energy region where the Roper 
contribution becomes dominant.  This feature allows the Roper 
contribution to be identified even without a clear corresponding peak in 
the cross section.  It is interesting to note the very different $D_{nn}$
behavior in Fig. 7 for $\theta_1 = 60^\circ$, where the Roper process 
provides a minor contribution over the entire $T_{\alpha'}$ range.  

Figure 8 shows results for $T_p = 2 ~GeV$ and $\theta_1 = 20^\circ $.  
Here, Roper excitation is manifested clearly in both the cross sections and  
$D_{nn}$.  Figure 9 reveals the real utility of the $D_{nn}$ signature, 
unveiling a Roper contribution at relatively high $T_{\alpha'}$,
where the net cross section is smooth and monotonically decreasing.

In Figs. 10 and 11, we show the results for a much higher energy,  
$T_p = 10 ~GeV$.  In these figures we can see that $D_{nn}$ reaches a 
maximum value around the peak of the Roper contribution, before decreasing 
toward higher $T_{\alpha'} $, because the $\Delta$ contribution has a 
longer tail in the cross section than the Roper contribution. 
In such situations, $D_{nn}$ measurements may yield information on the
$T_{\alpha'}$-dependence of the contributing production processes
far from the regions where they are kinematically maximized.

\section{Summary}

We have studied Roper resonance excitation in both 
the inclusive $ p + \alpha \rightarrow \alpha 
+ X$ reaction and the exclusive $\vec{p}+\alpha \rightarrow \vec{p}+\alpha+X$ 
reactions at $T_p = 1 - 10 ~GeV$.  
We have used a reaction model developed previously to 
understand existing inclusive cross section measurements. The model
includes the $\Delta$ excitation 
process in the $\alpha$-particle as well as the Roper excitation process 
in the proton.  We have calculated the differential cross 
sections and the normal spin transfer coefficient $D_{nn}$ for various energies 
and angles of the recoil $\alpha$-particle.  

The inclusive reaction sometimes exhibits a peak from the Roper resonance 
excitation.  The magnitude of the cross section does not have a strong 
dependence on the recoil 
$\alpha$ angle, in contrast to the case with inverse kinematics, since the 
momentum transfer to the $\alpha$-particle does not depend on its recoil 
angle.  Instead, the shape and strength of the Roper contribution to the
inclusive spectrum depend on the recoil 
$\alpha$ angle because of its kinematic implications for the invariant 
mass of the final $\pi N$ system. 

In the exclusive $\vec{p}+\alpha \rightarrow \vec{p}+\alpha+X$ 
reactions, we have calculated both the cross section and the spin transfer 
coefficient.  The simple spin coupling for Roper production dictates
that the incident proton's polarization normal to the production plane
will be transferred completely to the N$^*$.  
In the restricted part of phase space described in 
section 3, we have consequently found that the spin transfer coefficient clearly 
shows the contribution from the Roper excitation process even when there is 
no corresponding peak structure in the cross section.  By observing 
$D_{nn}$, one can distinguish the 
Roper process from the $\Delta$ background even when the energy spectrum is 
rather flat.  We conclude that the spin transfer coefficient is a robust 
observable for identifying the Roper contribution.  

If the polarization transfer measurements proposed here were to
confirm the dominance of Roper excitation in p+$\alpha$ collisions 
under appropriate kinematic conditions, then coincidence experiments 
with polarized beam offer several potential advantages over other 
methods for determining so far rather poorly known properties of the 
Roper resonance.  By changing the proton bombarding energy and the 
$\alpha$-particle recoil angle, one can vary the invariant mass of 
the excited nucleon independently of the momentum transfer to the 
$\alpha$-particle.  In this way, one can measure the resonance shape 
and improve upon existing determinations of its mass and width.  
Furthermore, the {\em a priori} knowledge of the N$^*$ polarization 
will help to determine the relative branching ratios for decay 
channels other than $\pi$N.  For example, by gating on p+$\alpha$
missing mass one could selectively study the N$\pi\pi$ channels,
which are known to have substantial contributions from $\Delta \pi$,
N$\rho$ and N$(\pi\pi)_{s-wave}$ intermediate states.  The different
intermediate states have different spin coupling, hence, different
characteristic spin transfers from N$^*$ to daughter N.  Measurement
of the polarization transfer from incident to final proton, as a
function of the reconstructed emission angle in the N$^*$ rest frame,
could then allow an improved decomposition of the $\pi\pi$ channel
strength.  The coupling strength of the N$^*$ to these various
channels is essential information for constraining theoretical models
of the Roper's structure.

At the higher bombarding energies considered here, it is of course also
possible to produce heavier baryon resonances, which have not been included
in the present calculation.  The cross sections for such production processes
may also be sizable, since for low $\alpha$-particle kinetic energies, the 
$\alpha$-particle form factor does not suppress the cross section.  
There are also
possibilities to use the same kind of spin filtering for certain heavier
resonances as applied to the Roper resonance in the present case. In 
particular, similar parity 
constraints on $D_{nn}$ to N$^*$ resonances, produced in 
exclusive $p + \alpha$ reactions, exist whenever the spin and parity of the 
resonances is $\frac{1}{2}^+$ ($D_{nn}=+1$) or $\frac{1}{2}^-$ 
($D_{nn}=-1$).  Furthermore, the full polarization transfer to the 
daughter baryon, when it is emitted along the resonance's spin 
quantization axis, applies equally well to $p + \pi$, $p + \eta$, and 
$\Lambda + K$ final states.  In the latter case [relevant, for example,
in the N$^*$(1710) decay], the polarization of the daughter baryon can
be readily measured via the $\Lambda$'s subsequent self-analyzing decay
to $p \pi^-$.
Thus, polarization transfer measurements in 
multi-GeV $p + \alpha$ collisions may help in the search for 
$\frac{1}{2}^+$ and $\frac{1}{2}^-$ strength in the nucleon resonance 
continuum.  \\
 
\noindent
{\bf Acknowledgment}: We acknowledge fruitful discussions with 
Prof. J. T. Londergan.  We also acknowledge Prof. L. Bland for 
useful discussions regarding the design of experimental apparatus to 
pursue the exclusive Roper measurements.  
We thank Prof. H. Toki for many useful suggestions.  
One of us (S.H.) acknowledges 
many discussions and collaborative works on the Roper resonance with Prof. E. 
Oset, and also the hospitality of IUCF and Indiana University, 
where this work was carried out.

\pagebreak

\pagebreak

\noindent
Fig.1 

\noindent
Diagrams for the $\vec{p}+\alpha \rightarrow \vec{p}+\alpha+X$ 
reactions considered in this paper.  They are: (a) the $\Delta$ excitation in 
the $\alpha$ \cite{Pedro95} and (b) the Roper excitation in the proton \cite{Hirenzaki96}. 
The $\sigma$ exchange must be interpreted as an effective interaction in 
the isoscalar exchange channel \cite{Hirenzaki96}. \\

\noindent
Fig. 2

\noindent 
Definitions of the kinematical variables used in this paper.  The scattering 
plane is 
determined by $\vec{p}$ and $\vec{p}_{\alpha'}$.  As indicated in this 
figure, $\vec{p}_{\pi N}$ is in  
the plane, while $\vec{p'}$ and $\vec{p}_{\pi}$ can be out of the plane.  
The incident proton polarization is perpendicular to the plane.  Definitions 
of the scattering angles are also shown. \\

\noindent
Fig. 3

\noindent
Calculated energy spectrum (solid line) and contribution from Roper 
excitation process alone (dashed line) for the inclusive 
$ p + \alpha \rightarrow \alpha + X$ reaction at $T_p = 1 ~GeV$ as a 
function of recoil $\alpha$-particle energy 
$T_{\alpha'}$.  The recoil $\alpha$ angles (in degrees) in the laboratory 
frame, correspond to values of $\theta_1$ defined in Fig. 2, 
and are indicated for each spectrum.  \\

\noindent
Fig. 4

\noindent 
Same as Fig. 3 except for  $T_p = 10 ~GeV$. \\

\noindent
Fig. 5

\noindent
Final proton emission angles (a) $ cos( \theta_2 )$, (b) $cos 
(\theta_3)$, and (c) final proton kinetic energies as a function of the 
final $\alpha$-particle  
kinetic energy $T_{\alpha'}$ for $T_p = 2 ~GeV$.  The curves correspond to 
different scattering angles of the $\alpha$, $\theta_1$, 
in the laboratory frame in units of degrees.  See Fig. 2 for a definition
of the scattering angles. \\ 

\noindent
Fig. 6

\noindent
(a) Differential cross section and (b) spin transfer coefficient 
$D_{nn}$ of the $\vec{p}+\alpha \rightarrow \vec{p}+\alpha+X$ reaction 
as a function of the recoil $\alpha$-particle kinetic energy 
$T_{\alpha'}$ at $T_p = 1 ~GeV$ and $\theta_1 = ~20^\circ$ . The dashed, 
solid, and thick solid lines show the results of the $\Delta$ process, 
the Roper process, and the combination of the two (including the 
interference). \\

\noindent
Fig. 7

\noindent
Same as Fig. 6 except for  $T_p = 1 ~GeV$ and $\theta_1 = 60^\circ$. \\

\noindent
Fig. 8

\noindent
Same as Fig. 6 except for  $T_p = 2 ~GeV$ and $\theta_1 = 20^\circ$. \\

\noindent
Fig. 9

\noindent
Same as Fig. 6 except for  $T_p = 2 ~GeV$ and $\theta_1=60^\circ$. \\

\noindent
Fig. 10

\noindent
Same as Fig. 6 except for  $T_p = 10 ~GeV$ and $\theta_1 = 20^\circ$. \\

\noindent
Fig. 11

\noindent
Same as Fig. 6 except for  $T_p = 10 ~GeV$ and $\theta_1 = 60^\circ$. \\

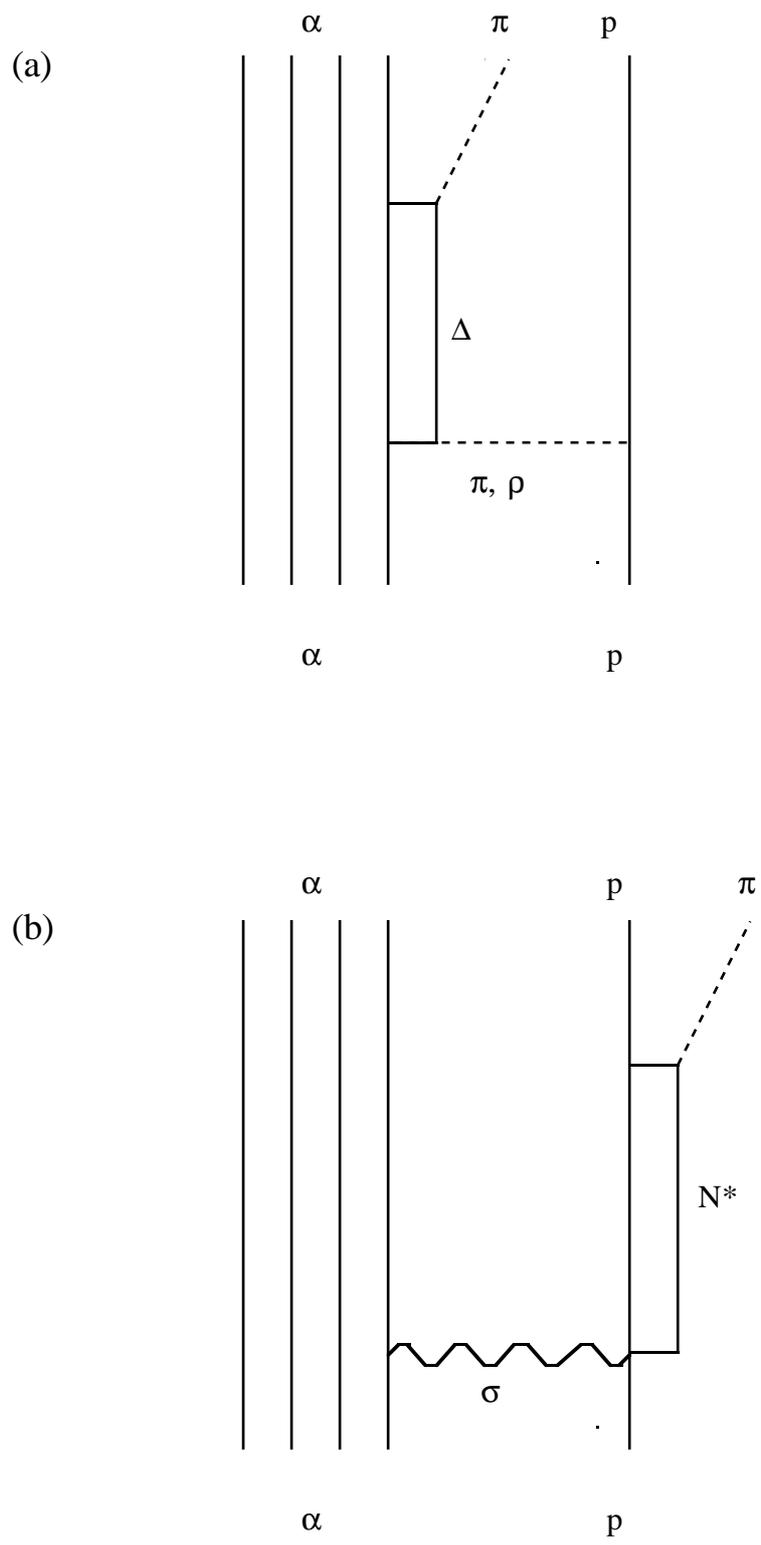
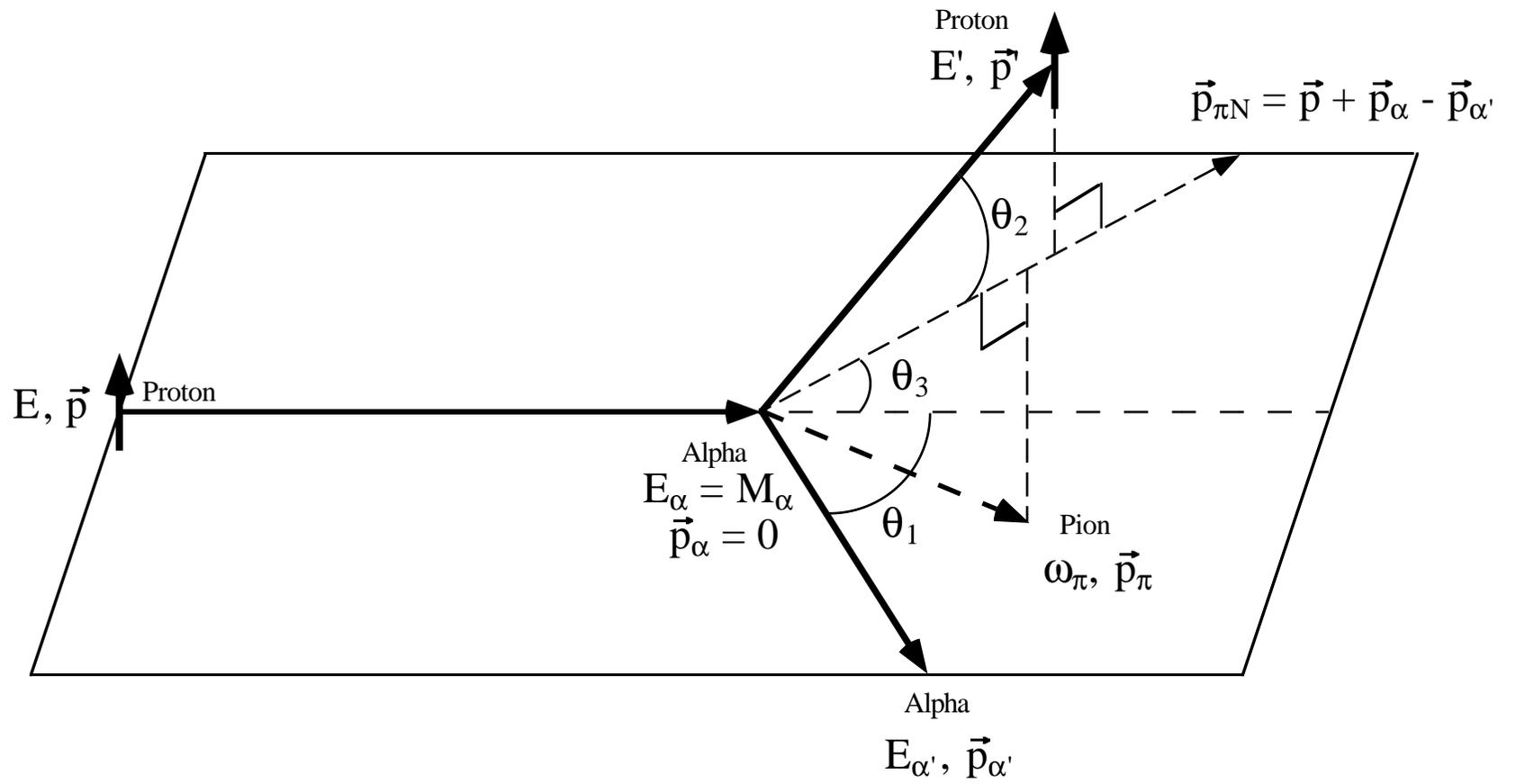
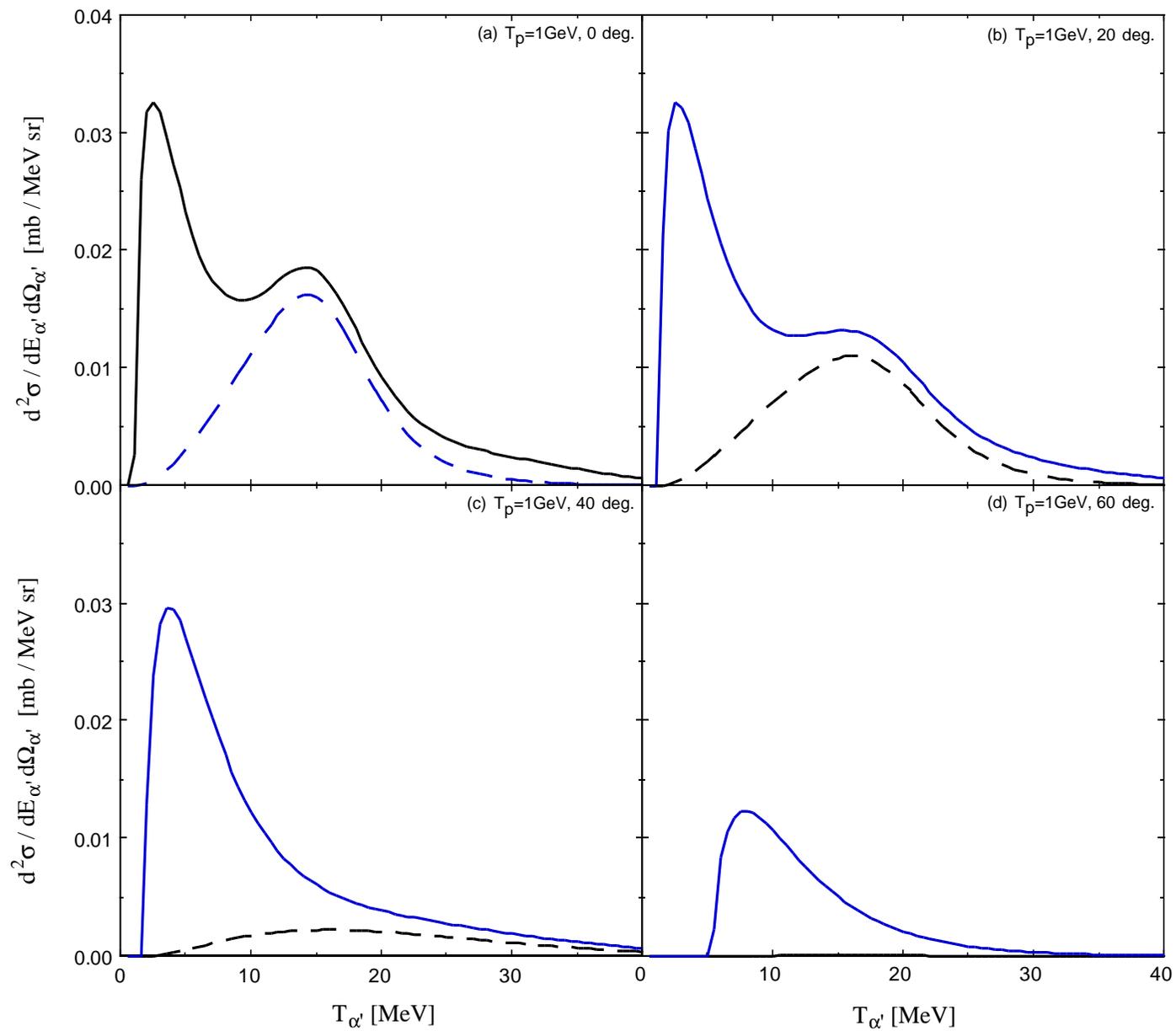
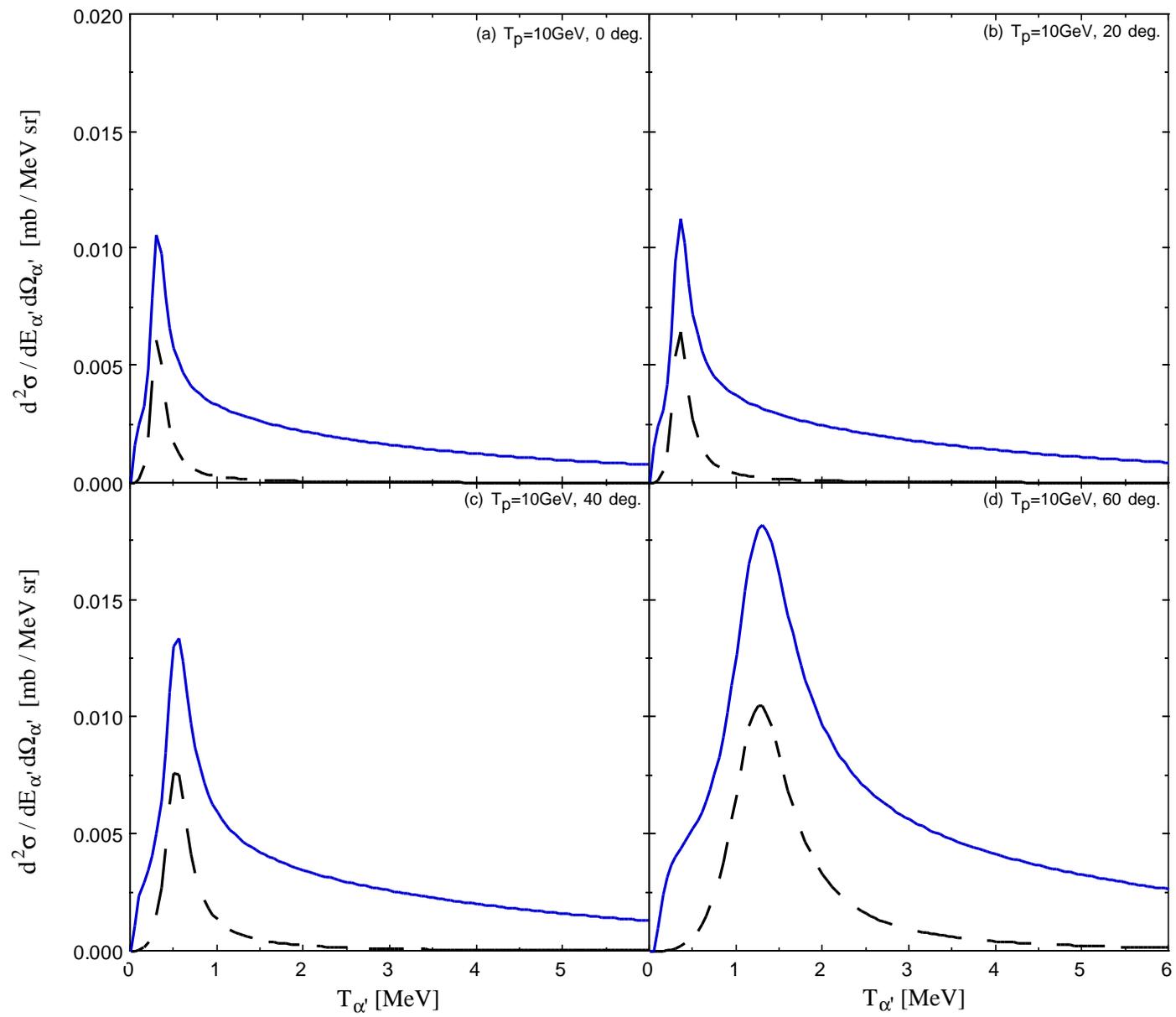
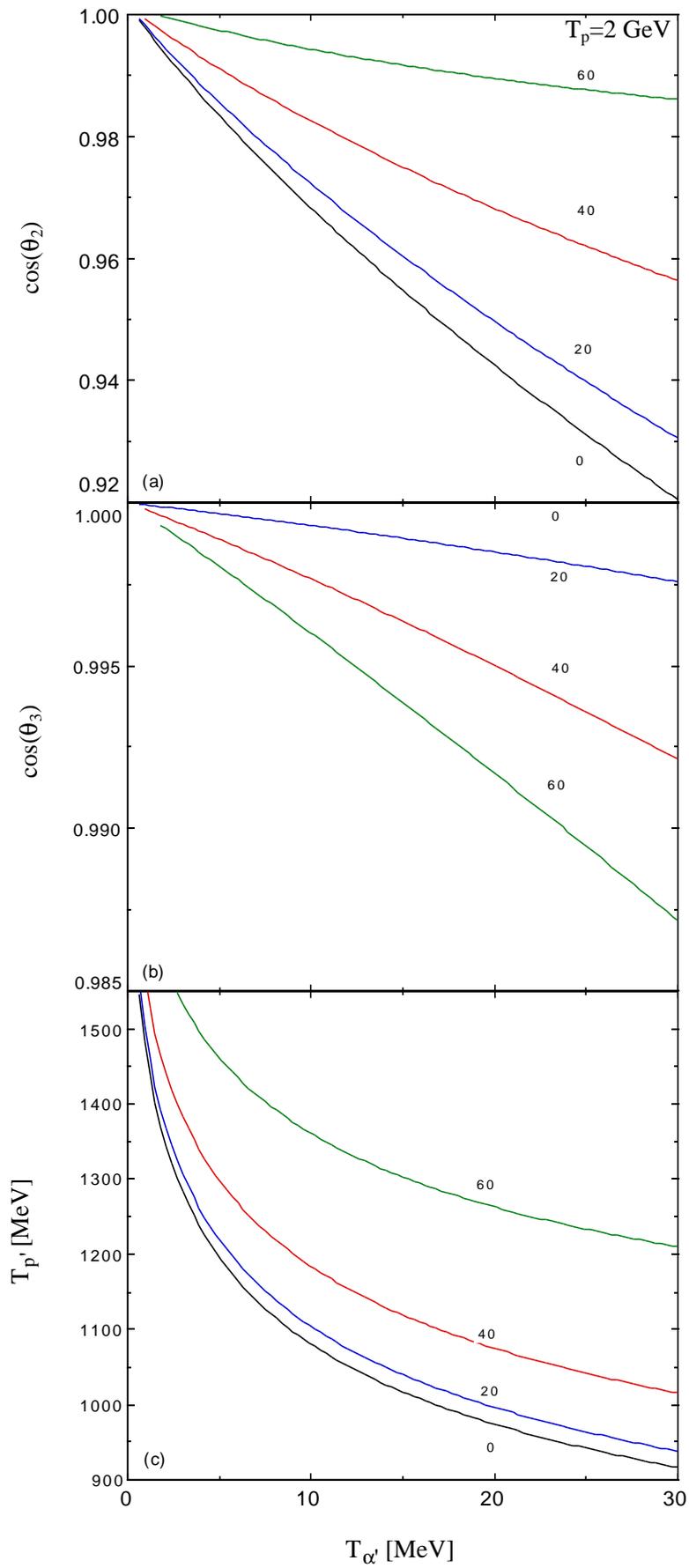
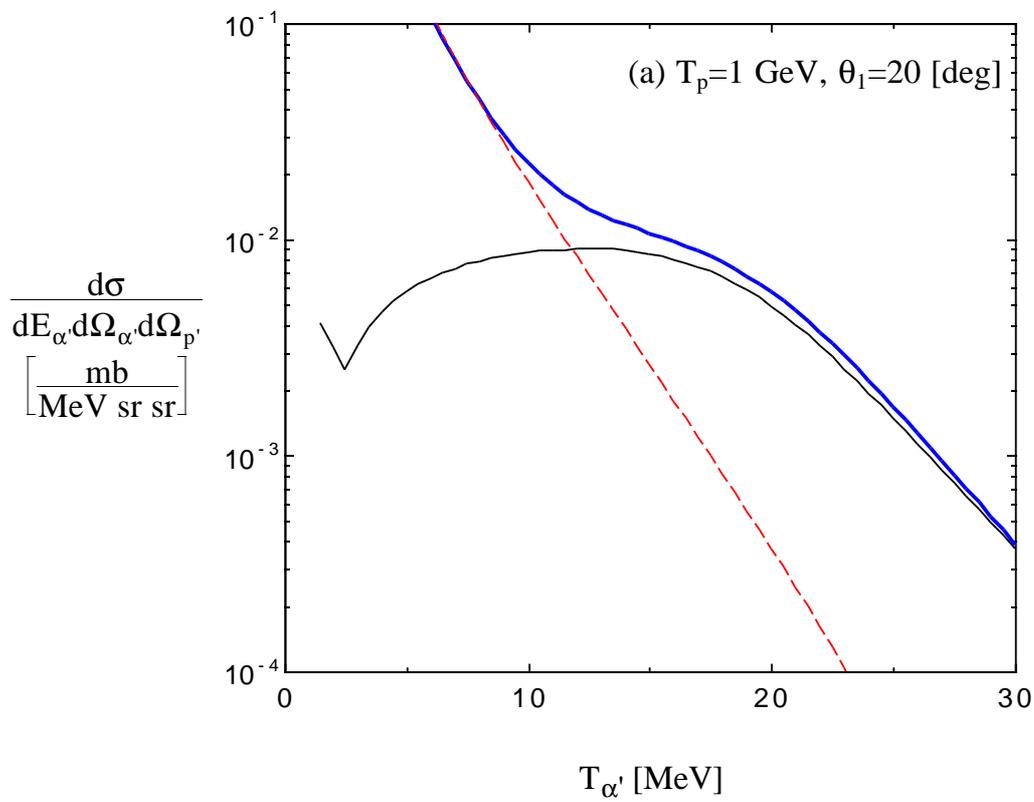
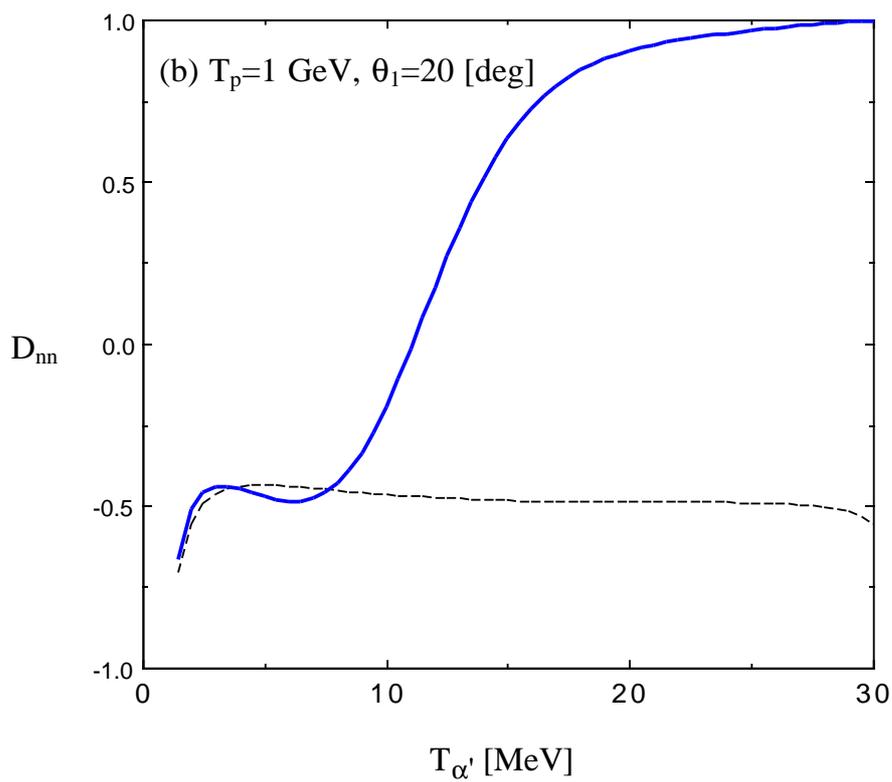
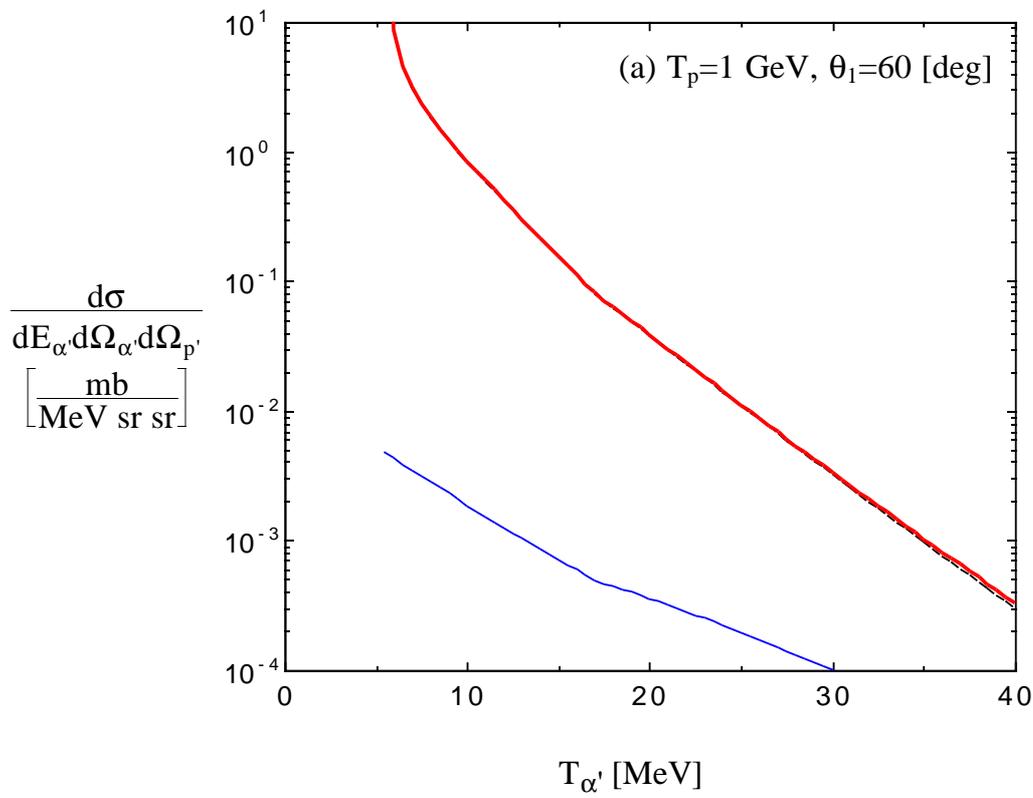
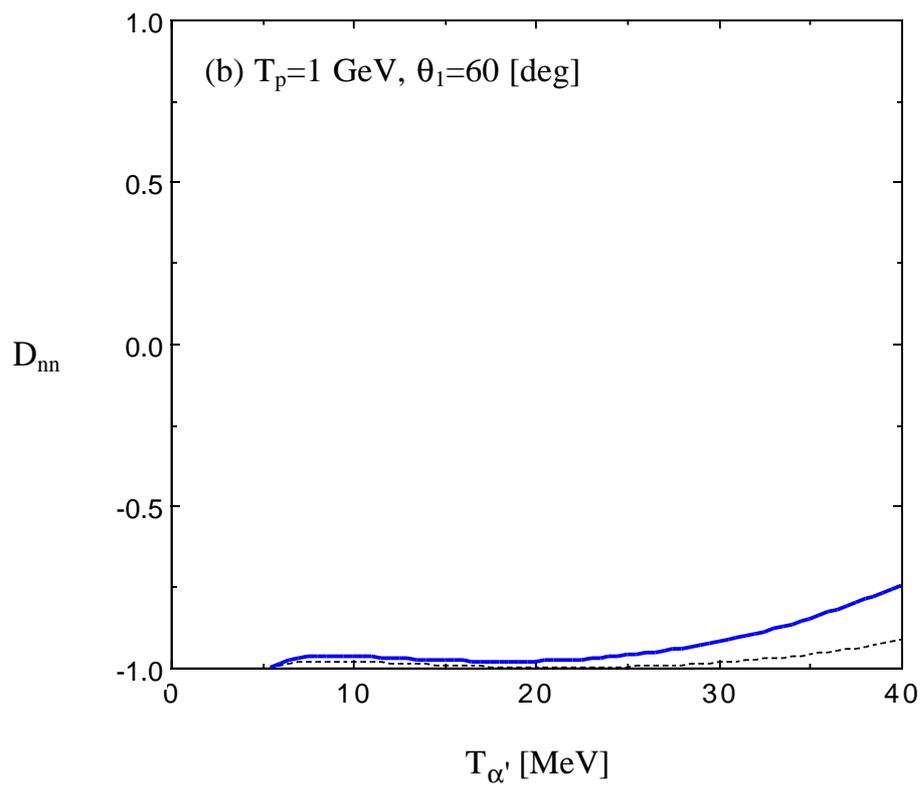
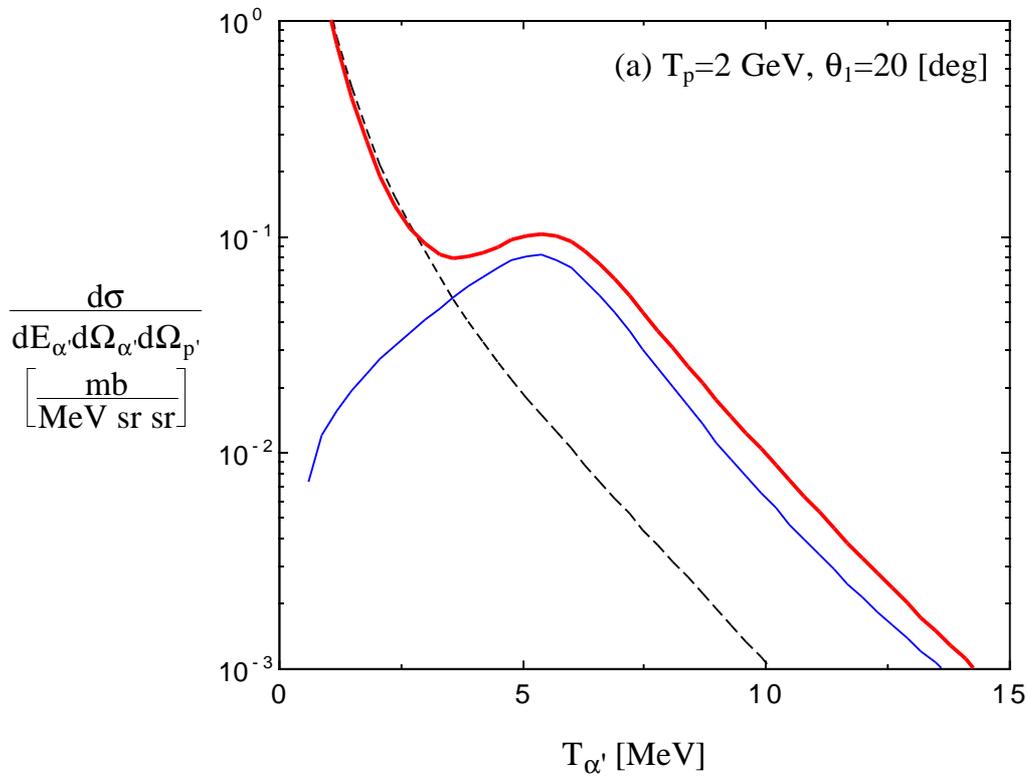
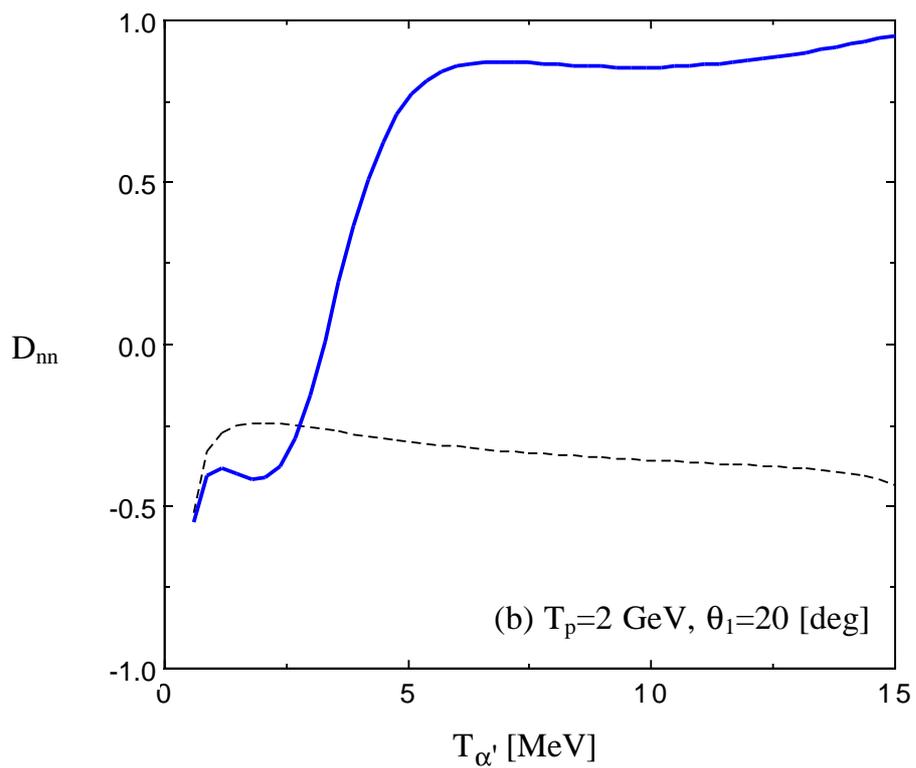
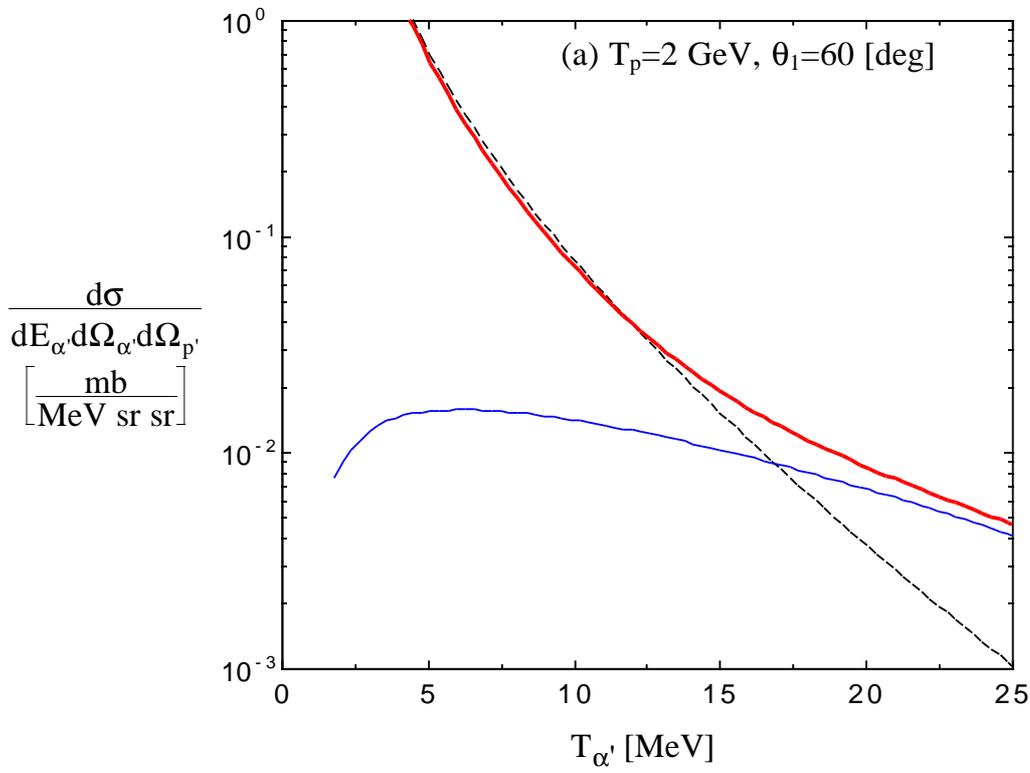
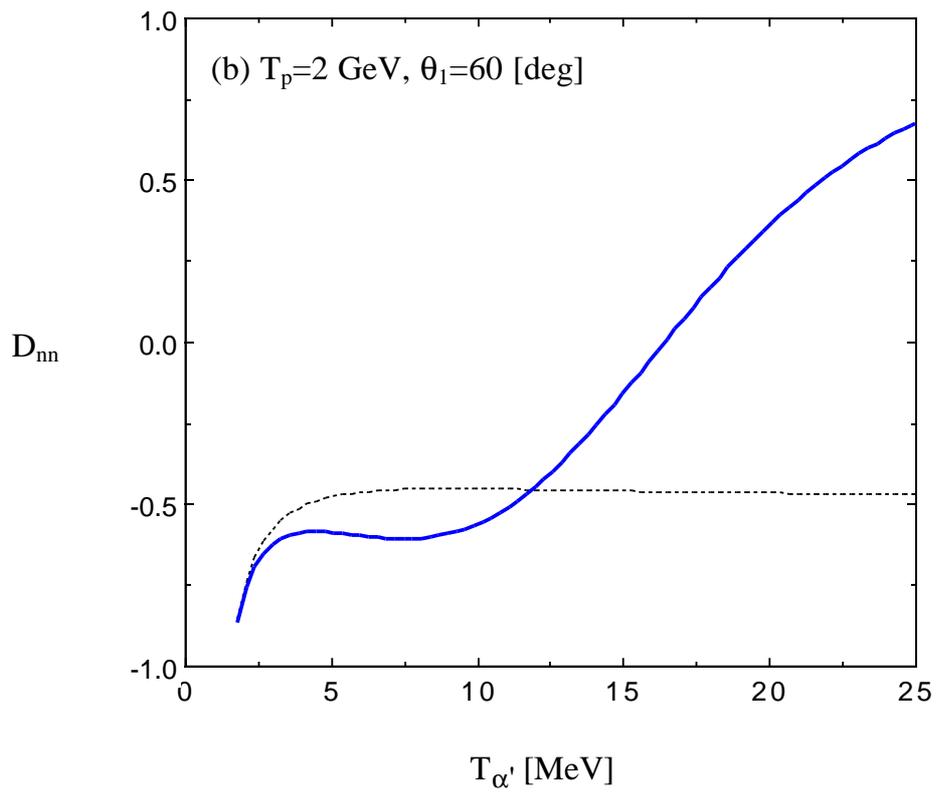
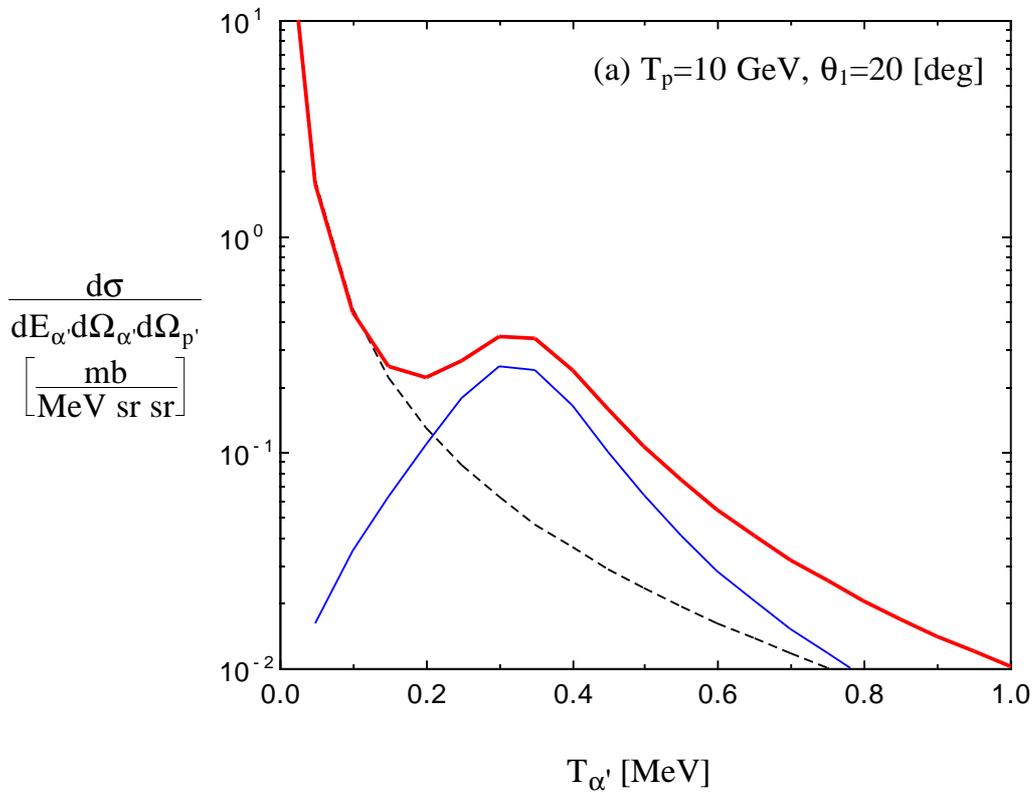
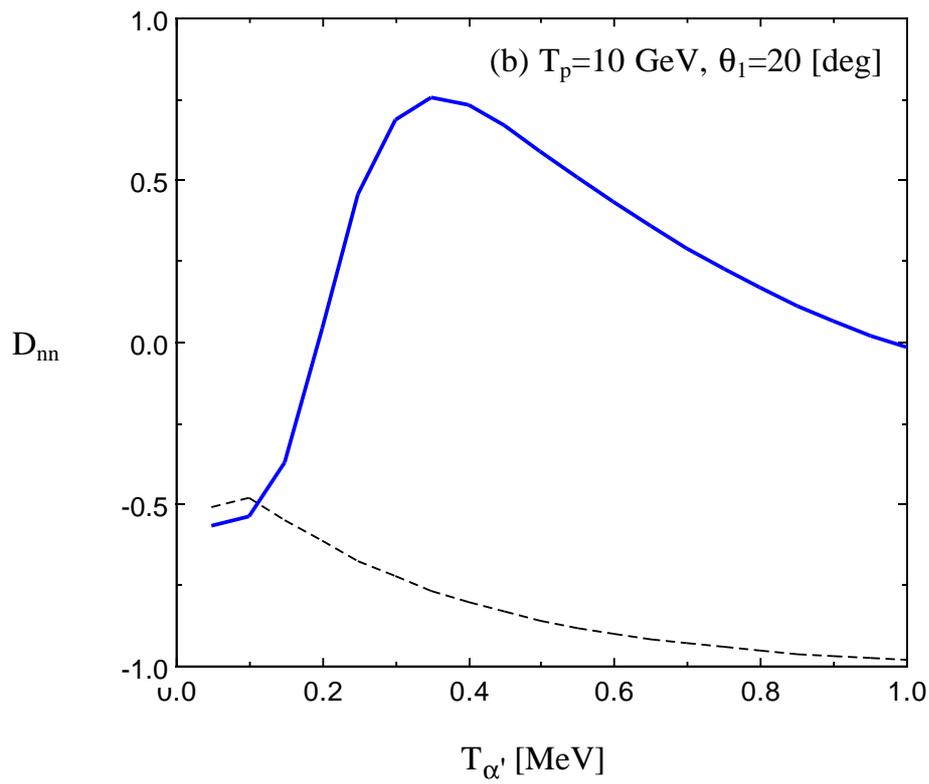
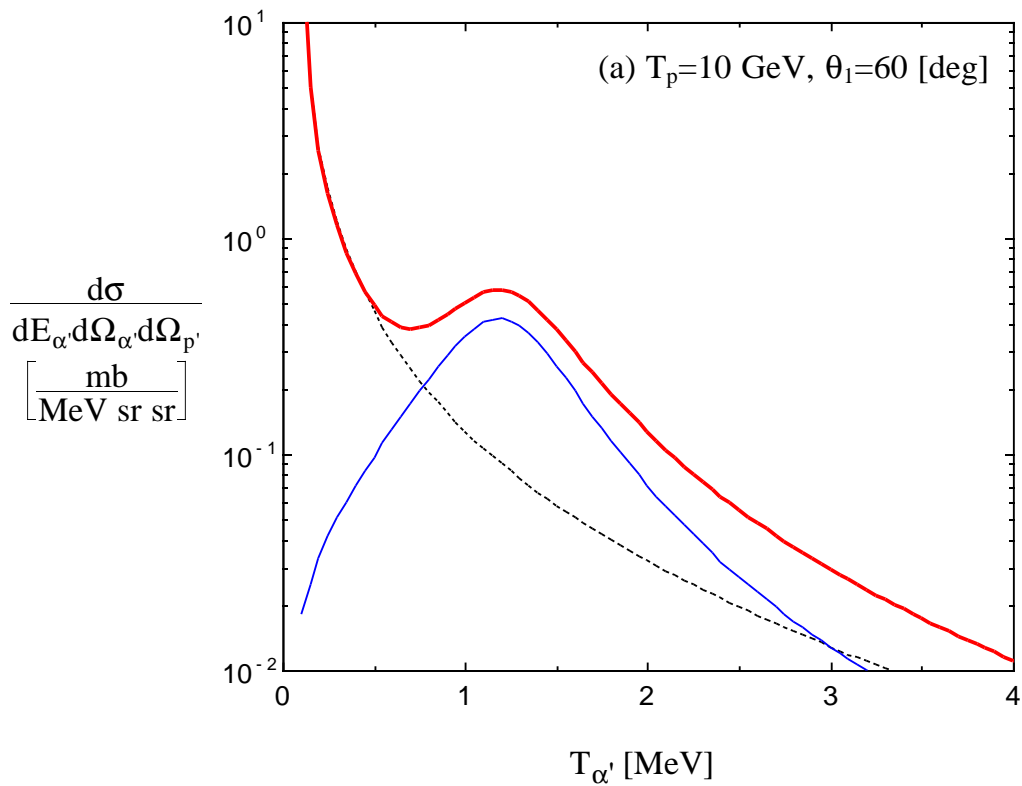
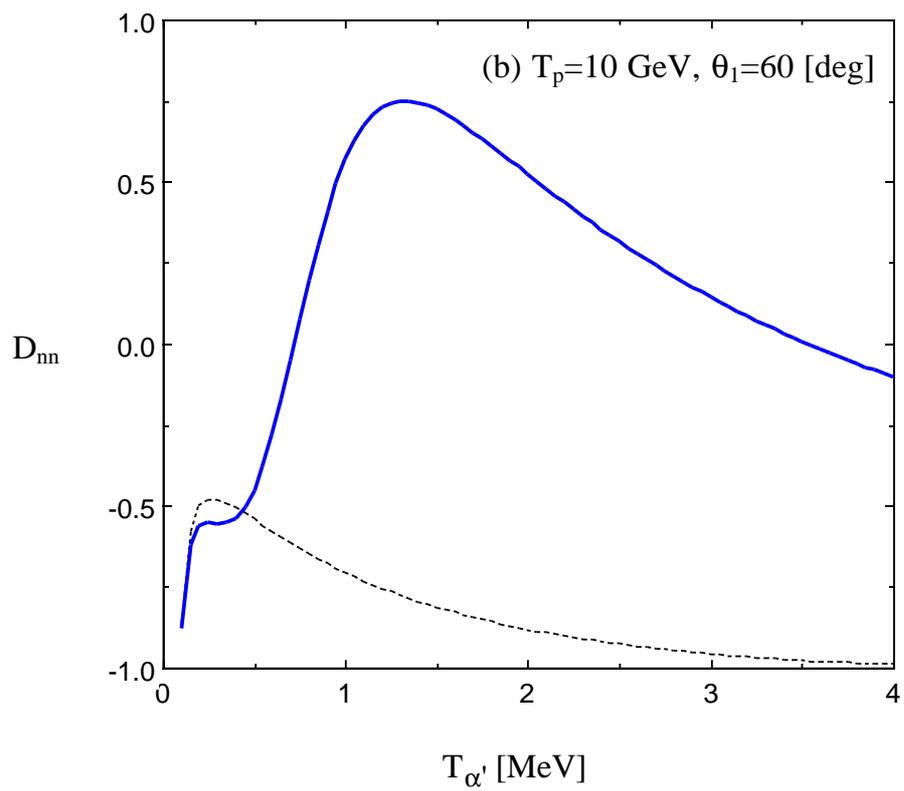
\end{document}